\documentclass[%
 reprint,
 amsmath,amssymb,
 aps,
]{revtex4-2}
\usepackage{physics}
\usepackage{epsfig}
\usepackage{graphicx}
\usepackage{dcolumn}
\usepackage{bm}
\usepackage{mathtools}
\usepackage{natbib}

\begin{document}

\preprint{APS/1-DFIST}

\title{Extended Field Interactions in Poisson's Equation Revision}


\author{Mario J. Pinheiro}
\affiliation{%
 Instituto Superior T\'ecnico - IST, University of Lisbon, Department of Physics, Av. Rovisco Pais, 1049-001 Lisboa Codex, Portugal.}%




\date{\today}

\begin{abstract}
In the present research, a variational technique to modifying the Poisson equation is presented, expanding its modelling capabilities to include a wider range of physical processes and resonant structures. The study examines the implications of this modified equation to further develop our knowledge of electrostatic potentials and providing a nonlocal extension of Einstein's theory of gravitation, energy conversion and communication, in addition to the methodological advancements. 
In addition to providing insights into the dynamics of gravitational potential in systems exhibiting radial vorticity fluctuation, the paper clarifies nonlocal effects, examines longitudinal electromagnetic wave generation, and resonant phenomena in dusty plasma media.

\end{abstract}

\maketitle


\section{Introduction}

The Poisson equation, fundamental in mathematical physics, provides critical insights across various domains, from electrostatics to gravitational fields~\cite{Poisson, Maxwell}. 
We propose a novel formulation of the Poisson equation in this work, which is based on a variational approach~\cite{Pinheiro2013}. This new method makes it possible to balance entropy maximisation with energy minimization, producing differential equations that provide an expanded understanding of physical systems. Beyond the conventional range, our updated equation provides critical information for studying resonant structures in electromagnetic, heat transmission, and acoustic systems \cite{Myers1998, Wright2001, Zutic2004, Serdyukov2001}. As a foundational tool in various scientific domains, this adaptable second-order partial differential equation is typically solved by a range of analytical and numerical approaches (see, e.g., Refs.~\cite{Gilbarg1983,Reddy2006,Mammoli2018,Trefethen1997}).

This work highlights the modified Poisson equation's significance in comprehending complex physical events by examining its numerous applications. We discuss the implications for nonlocal gravity theories~\cite{Tarassov2022,Mashhoon2022}, spintronics~\cite{Duadi,Lu,Dei}, and resonant structures~\cite{Jonatan,Ludovic,Joseph}, demonstrating how it might improve our understanding of complex structures with potential significance to technological and scientific domains.

Here is an overview of the manuscript: In Section 2 builds upon our prior research introducing the concept of topological torsion current in non-equilibrium physical systems \cite{Pinheiro2016, Pinheiro2022}. We focus on the role of this concept, elucidated through variational methods, and its implications on the dynamics of the electric field, incorporating a novel geometric structure, \(\pmb{\omega}\). In Section 3, we investigate the diverse applications of the modified Poisson equation. The section covers its impact on understanding Yukawa's potential in atomic physics, detailing the inter-galactic gravitational potential, and exploring the modifications in spintronic current density. These discussions highlight the equation's adaptability and its capacity to model complex physical phenomena. In Section 4, we conclude by summarizing the key findings of our study and propose directions for future research. 

\section{Investigating Non-Equilibrium Systems: A Variational Approach to Equilibrium Dynamics}

In previous works \cite{Pinheiro2016,Pinheiro2022}, we introduced the concept of topological torsion current and its significant implications in spacecraft dynamics and particle physics. Building upon this foundation, we now explore its role in non-equilibrium physical systems, particularly through variational methods. Central to our approach is the introduction of a novel geometric structure represented by \(\pmb{\omega}\), which plays a pivotal role in our theoretical framework.

The electric field in such a system can be described by:
\begin{equation}\label{eq1}
\vb{E} = -\grad{\Phi} - \frac{\partial \vb{A}}{\partial t} - \widetilde{\pmb{\omega}}\; \Phi
\end{equation}
Here, the term \(\widetilde{\pmb{\omega}}\) is defined as:
\begin{equation}\label{eq1a}
\widetilde{\pmb{\omega}} = \frac{\pmb{\omega} \times [\vb{v}]}{c^2}
\end{equation}
In this formulation, \(\Phi\) represents the electric potential, \(\vb{A}\) is the vector potential, and \(\pmb{\omega}\), embodying the newly introduced geometric structure, signifies the angular velocity. This angular velocity component is not just a physical quantity but also encapsulates a spin connection, indicating the presence of a torsion field within the theoretical construct. We thus conceptualise the cross product $[\; \pmb{\omega} \times \vb{A}]$ as more than a simple mathematical operation, and instead comprehend it as the expression of a torsion field inside the electromagnetic field structure. Even though the symbol $\pmb{\omega}$ represents angular velocity, it also has a geometrical meaning that goes beyond its conventional physical meaning. This geometrical component, which is essential to our new methodology, presents torsion as an essential attribute of the field. In this situation, torsion is a physical representation of the inherent geometrical features of the field rather than just a mathematical abstraction. Broadening the conventional comprehension of electromagnetic phenomena by incorporating the concept of torsion, open up novel possibilities for investigating the interaction between geometric structures and physical forces.

Thus, when a charged particle \(q\) is taken into consideration, its Liénard-Wiechert potentials can be expressed as follows: \(\phi (\vb{r},t)=\frac{q}{4 \pi \epsilon_0}\frac{1}{(r-\vb{r} \vdot \vb{v}/c)}\). Here, \(\vb{r}\) is the vector from the retarded position to the field point \(\vb{r}\), and \(\vb{v}\) is the velocity of the charge at the retarded time. Thus, we express \(\vb{A}(\vb{r},t)\) as \(\frac{\vb{v}}{c^2}\phi(\vb{r},t)\). Consequently, \([\; \pmb{\omega} \cp \vb{A}]=\pmb{\widetilde{\omega}}\phi\).

By applying these concepts to Poisson's equation, which characterizes the electrostatic potential due to a charge distribution as \(\div{\vb{E}}=\frac{\rho}{\epsilon_0}\), we derive a new form of Poisson's equation. This formulation not only incorporates the standard electrostatic potential but also includes the influence of the novel geometric structure encapsulated in $\pmb{\widetilde{\omega}}$. This advancement opens the door to exploring resonant phenomena, particularly those observed in dusty-plasma mediums, through a lens that integrates both physical and geometric insights. Hence, we have:
\begin{equation}\label{eq2}
\div{\vb{E}}=-\laplacian{\Phi} - \partial_t \div{\vb{A}} - \div{\pmb{\widetilde{\omega}}\; \Phi}.
\end{equation}
We remind that the Coulomb gauge constrains $\div{\vb{A}}=0$, and the Lorenz gauge, $\div{\vb{A}}=-\partial_t \Phi$. Hence, in the Lorenz gauge Laplace equation must be written the inhomogeneous wave equation under the new form

\begin{equation}\label{eq3}
\laplacian{\Phi} - \partial^2_{tt} \Phi=-\frac{\rho}{\epsilon_0} - (\div{\pmb{\widetilde{\omega}}}) \Phi - (\grad{\Phi} \vdot \pmb{\widetilde{\omega}}).
\end{equation}

Exploring the dynamics of charge density oscillations and Coriolis-like terms, we have non three distinct source terms: i) the charge density $\rho$ and the effects of its oscillations, the well-known plasma oscillations at high frequency and the magnetohydrodynamic waves at low frequency; ii) the divergence of the Coriolis-like source term, $\div{\widetilde{\omega}} \neq 0$; iii) a measure of the degree of alignment between the electric field and the Coriolis-like term, $(\grad{\Phi} \vdot \pmb{\widetilde{\omega}})$.

\section{Potential Applications}

Over the original equation, the revised version of the Poisson equation has a number of benefits. Since it was designed for spinning systems, it is broader and may therefore be used to mimic a wider range of physical systems, including galaxies and stars. Additionally, it is more precise and has a better ability to simulate resonant structures.
The new form of the problem may be solved using a variety of numerical techniques and is also computationally efficient. As a result, the equation may be solved fast and precisely, enabling the modeling of complicated physical systems in a relatively short amount of time.

\subsection{Some insight on Yukawa's potential}

The Yukawa potential, which is an important rotational-invariant scattering potential used in atomic and nuclear physics, can be derived from this formalism. By imposing the constraint $\div{\widetilde{\omega}} = 0$, we can retrieve this potential from the equation, and we can also obtain the Debye-H"{u}ckel equation for electrolytes. When considering a solenoidal vortex field with $\div{\widetilde{\omega}} = 0$, and in the stationary case with $\grad{\Phi} \vdot \pmb{\widetilde{\omega}}=0$, the well-known Yukawa equation for a particle of charge $\rho=Ze$ can be derived.

\begin{equation}\label{eq4}
\laplacian{\Phi} + \kappa^2 \Phi = -\frac{Ze}{\epsilon_0} \delta(\vb{r}-\vb{r'})
\end{equation}
Here, $\kappa = \frac{\pmb{\widetilde{\omega}}}{c^2}$ and the solution is

\begin{equation}\label{eq5}
G_{\omega}(\vb{r},\vb{r'}) = \frac{Ze}{4 \pi \epsilon_0} \frac{e^{\imath \kappa \abs{\vb{r} - \vb{r'}}}}{\abs{\vb{r} - \vb{r'}}}.
\end{equation}

If we assume a radially-dependent electric potential, Eq.~\ref{eq4} is reduced to
\begin{equation}\label{eq6}
\begin{aligned}
\partial^2_{rr} \Phi + \frac{2}{r}\partial_r \Phi + \left[ \frac{1}{r^2}\partial_r (r^2 \widetilde{\omega}_r) \right. & \\
\left. + \frac{1}{r \sin \phi}\partial_{\theta}\; \widetilde{\omega}_{\theta} \right] \Phi + \widetilde{\omega}_r \partial_r \Phi &= - \frac{Ze}{\epsilon_0}
\end{aligned}
\end{equation}

Consider simplifying to basic radial dependencies, such as $\widetilde{\omega}=\widetilde{\omega}(r)$. Then

\begin{equation}\label{eq7}
\partial^2_{rr} \Phi + \left( \frac{2}{r} + \widetilde{\omega}_ r \right) \partial_r \Phi + \left[ \frac{1}{r^2}\partial_r (r^2 \widetilde{\omega}_r) \right] \Phi = - \frac{Ze}{\epsilon_0}
\end{equation}

\begin{figure}
  \centering
  \includegraphics[width=3.5 in]{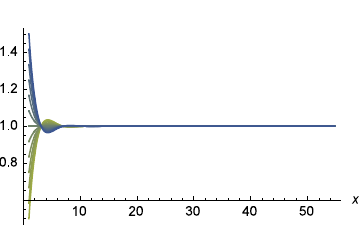}\\
  \caption{Family of potential curves as a function of distance $r$. Arbitrary units.}\label{Fig1}
\end{figure}

For the particular case $\widetilde{\omega}=1$, Fig.~\ref{Fig1} shows the analytical solution of Eq.~\ref{eq7}. Consider the fact that the potential might be either attractive or repulsive.

Analytical solutions to Eq.~\ref{eq7} are essential in understanding the behavior of physical systems. Eq.~\ref{eq7} is a special case of the general equation and its analytical solution is represented in Fig.~\ref{Fig1}. The potential can be of either the repulsive or attractive kind.

It is very similar to the Lennard-Jones potential. The Lennard-Jones potential is a mathematical model used to describe the interactions between atoms or molecules. It is a simple, effective model for describing the weak intermolecular forces that occur in nature. The Lennard-Jones potential is a combination of two terms: a repulsive term and an attractive term. The repulsive term is a function of the distance between the atoms and the attractive term is a function of the inverse of the distance between the atoms. However, while bearing similarity to the Lennard-Jones potential, introduces a novel aspect by incorporating the radial dependence of $\widetilde{\omega}(r)$. This inclusion differentiates it from the standard Lennard-Jones model, which typically focuses on the interplay of repulsive and attractive forces without a specific radial dependence component. 

The study of Yukawa potentials in dusty plasmas has significant implications for real-world applications. For instance, offers critical insights into the behavior of plasma under magnetic influences, crucial for advancements in controlled nuclear fusion technology~\cite{dynamics_and_transport}. Additionally, contribute to our understanding of plasma thermodynamics, potentially aiding in the development of more efficient energy systems~\cite{isomorph_invariance}, or sheds light on phase transitions in plasma, a key aspect that could drive innovations in materials science~\cite{molecular_dynamics_study} .

\subsection{Inter-galactic Gravitational potential}

The behavior of gravitational potential in a system where the radial component of the vorticity varies inversely with distance from the center is explored in this Section. The assumption that the radial component is proportional to $1/r$ leads to a solution of the potential in terms of three terms: a term proportional to $1/r$, a term proportional to $\ln(r)/r$, and a term proportional to $r^2$. This result sheds light on the behavior of gravitational potential in idealized and real fluids and is consistent with theoretical predictions and astronomical observations. The obtained intergalactic gravitational potential is consistent with the one caused by dark matter and presents the Navarro-Frenk-White (NFW) profile equation for describing the density distribution of dark matter halos in cosmological simulations.

If the source has a gravitational origin, Eq.~\ref{eq7} must be written under the form

\begin{equation}\label{eq8}
\partial^2_{rr} \Phi + \left( \frac{2}{r} + \widetilde{\omega}_ r \right) \partial_r \Phi + \left[ \frac{1}{r^2}\partial_r (r^2 \widetilde{\omega}_r) \right] \Phi = - 4 \pi G \rho.
\end{equation}

\begin{figure}
  \centering
  \includegraphics[width=3.5 in]{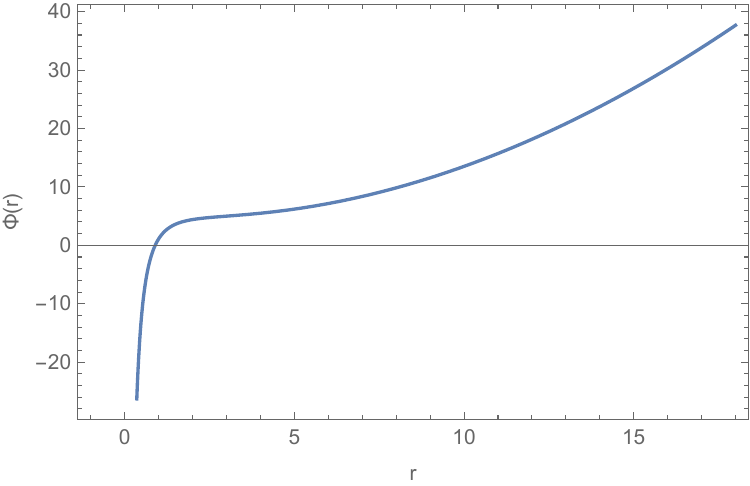}\\
  \caption{Inter-galactic gravitational potential \(\Phi(r)\), as a function of the radial distance \(r\) (in arbitrary units).}\label{Fig2}
\end{figure}

If we assume $\widetilde{\omega}_r=1/r$, then it follows the solution of the potential
\begin{equation}\label{eq9}
\Phi(r)=\frac{c_1}{r}+\frac{c_2 \ln(r)}{r}+\frac{r^2}{9}
\end{equation}
and its representation in Fig.~\ref{Fig2}.

From this result, we can conclude that the gravitational potential $\Phi(r)$ can be expressed as a combination of three terms: the classical term proportional to $1/r$, a term proportional to $\ln(r)/r$, and a term proportional to $r^2$. The coefficients of these terms are given by $c_1$, $c_2$, and $1/9$, respectively. The assumption that $\widetilde{\omega}_r=1/r$ allowed us to solve for the potential $\Phi(r)$ using the equations governing the behavior of gravitational fields. This result provides insight into the behavior of the gravitational potential in a system where the radial component of the vorticity is assumed to vary inversely with the distance from the center of the system. A similar result was obtained in the work of Farkhat Zaripov~\cite{Zaripov} in the context of dark matter models, where the gravitational potential's characteristics were explored within the framework of modified gravity theories. Zaripov's study, like ours, highlights the complex interplay between field oscillations and gravitational effects, underscoring the versatility and depth of these theoretical models in astrophysical contexts.

This is consistent with the theoretical predictions and astronomical observations~\cite{Hayashi}. In an idealized inviscid fluid, the circulation of a vortex is conserved and does not decay with distance. In a real fluid with viscosity, the circulation can decay with distance due to viscosity effects. In 2D viscous flows, the decay rate of circulation with distance can be approximated as 1/r. In 3D flows, the decay rate can be more complex and can depend on the specific flow conditions and properties.

A modified version of the gravitational potential equation that accounts for the spread of dark matter in the cosmos may be used to determine the intergalactic gravitational potential caused by dark matter. A useful equation for describing the density distribution of dark matter halos in cosmological simulations is the Navarro-Frenk-White (NFW) profile. The source of the NFW profile is:

\begin{equation}
    \rho(r) = \frac{\rho_0}{(r/r_s)(1+r/r_s)^2}
\end{equation}

where $\rho(r)$ is the density of dark matter at a distance $r$ from the center of the halo, $\rho_0$ is a characteristic density, and $r_s$ is a scale radius.

Using the Poisson equation, we can relate the gravitational potential to the dark matter density as, $$\nabla^2\Phi = 4\pi G\rho$$, where $\Phi$ is the gravitational potential and $G$ is the gravitational constant.

Substituting the NFW profile into this equation and solving for $\Phi$, we obtain the following expression for the inter-galactic gravitational potential due to dark matter:
\begin{equation}~\label{eq12}
    \Phi(r) = -\frac{4\pi G\rho_0r_s^3}{r}\ln\left(1+\frac{r}{r_s}\right).
\end{equation}
The second term obtained in Eq.~\ref{eq9} bears resemblance to that of Eq.~\ref{eq12}.

\subsection{A modification of the Spintronic Current Density}

Spintronics is an emerging field of technology that focuses on the manipulation and control of electrons' spin in solid-state devices~\cite{intro_spintronics}. The spintronic current density can be optimized to increase performance while using less energy and to better integrate with cutting-edge technology like quantum computers and neural networks.
We get the total current density from Eq. ref. eq.1 and the general constitutive relation $\vb{J}=\sigma_c \vb{E}$.
\begin{equation}\label{eq10}
\vb{J}=\sigma_c \vb{E}_0 - \sigma_c \frac{\partial \vb{A}}{\partial t} - \sigma_c \; \pmb{\widetilde{\omega}} \;\Phi.
\end{equation}

In the context of spintronics, Eq.~\ref{eq10} suggests that the current density $\vb{J}$ can be decomposed into three terms: the first term $\sigma_c \vb{E}_0$ is the ordinary Ohmic current, the second term $-\sigma_c \frac{\partial \vb{A}}{\partial t}$ represents the contribution from the electromagnetic induction due to time-varying magnetic fields, and the third term $-\sigma_c \pmb{\widetilde{\omega}}\Phi$ represents the contribution from the spin accumulation induced by the spin-orbit interaction.

One possible manipulation of Eq.~\ref{eq10} is to take its curl, which can give us insights into the behavior of the spintronic current density under certain conditions. Taking the curl of both sides of Eq.~\ref{eq10}, we obtain:

\begin{equation}
\nabla \times \vb{J} = \sigma_c \nabla \times \vb{E}_0 - \sigma_c \frac{\partial}{\partial t} (\nabla \times \vb{A}) - \sigma_c \nabla \times (\pmb{\widetilde{\omega}}\;\Phi).
\end{equation}

Using the Maxwell-Faraday equation $\nabla \times \vb{E} = -\frac{\partial \vb{B}}{\partial t}$, we can simplify the second term on the right-hand side:

\begin{equation}
\nabla \times \vb{J} = \sigma_c \nabla \times \vb{E}_0 + \sigma_c \frac{\partial \vb{B}}{\partial t} - \sigma_c \nabla \times (\pmb{\widetilde{\omega}}\Phi).
\end{equation}

Assuming that the magnetic field $\vb{B}$ is negligible compared to the electric field $\vb{E}_0$, we can drop the second term and simplify further:

\begin{equation}
\nabla \times \vb{J} = \sigma_c \nabla \times \vb{E}_0 - \sigma_c \nabla \times (\pmb{\widetilde{\omega}}\Phi).
\end{equation}

This formula demonstrates how the axial vector term involving the spin $\pmb{\widetilde{\omega}}$ and the curl of the electric field relate to the curl of the spintronic current density. This connection may be leveraged to comprehend the dynamics of spin currents in intricate geometries and to develop new spintronic devices.

Examining the behaviour of the axial vector term $\pmb{\widetilde{\omega}}\Phi$ in Eq.~\ref{eq10} when other spin textures or geometries are present is one method that might be used. Consider, for instance, the situation of a magnetic domain wall in which the spin direction fluctuates from side to side. This may result in the spin current density having a non-zero value, which could have potential spintronics applications.
This leads to a deeper understanding of spintronic current density and its applications~\cite{Cheng,Qi,Angizi,Jonietz,Cockburn}.

\subsection{Applications to Nonlinear Optics and Metamaterials}

The ability to generate and control longitudinal EM waves is still in its inception. Longitudinal electromagnetic waves have the potential to revolutionize communication and data storage. They could be used to transmit data over long distances, with no loss of quality or speed. These waves could also be used to store vast amounts of data, making it possible to store and access large amounts of information in a relatively small space. Classical Maxwell's equations predict the existence of only transversal waves in a vacuum since without material support, such as in a plasma medium where Langmuir waves are observed, there is no support for their propagation. Theoretical evidence for the existence of longitudinal waves in a vacuum has been documented in several sources, including~\cite{Waser_2001,Vlaenderen,Khvorostenko_2002}, at observational level~\cite{Wesley_2002,Podgany_2011}, and ~\cite{Hively} proposal of two different apparatuses that are configured to transmit and/or receive scalar-longitudinal waves. Further experimental evidence is needed to confirm their existence (see Refs.~\cite{Jefimenko, Kong, Lindell, He, Shukla, Wu}). Attenuation is the gradual loss of energy that occurs as waves propagate through a medium. However, unlike transverse electromagnetic waves, longitudinal waves are not subject to attenuation effects since they can travel through a vacuum without a medium. This unique characteristic allows them to transmit data over long distances without any loss of quality or speed, which could revolutionize the field of communication. With the ability to transmit large amounts of data over vast distances, global communication, and information exchange could be greatly enhanced.

Upon calculating the gradient of Eq.~\ref{eq3} and substituting $
\Box u = \Delta u - \frac{1}{c^2} \frac{\partial^2 u}{\partial t^2}$, it becomes apparent that
\begin{displaymath}\label{eq11}
\square \vb{E}=\frac{\grad{\rho}}{\epsilon_0}+(\div{\widetilde{\pmb{\omega}}} )\vb{E}+\Phi (\grad{\div{\widetilde{\pmb{\omega}}}}) + (\vb{E} \vdot \grad)) \;\widetilde{\pmb{\omega}} +
\end{displaymath}
\begin{equation}~\label{eq11a}
(\widetilde{\pmb{\omega}} \vdot \grad) \vb{E} + \vb{E} \cross [\curl{\widetilde{\pmb{\omega}}}] + \widetilde{\pmb{\omega}} \cross [\curl{\vb{E}}].
\end{equation}
This equation describes the behavior of the electromagnetic field, specifically the electric field $\vb{E}$, in the presence of a source charge density $\rho$ and vorticity field $\pmb{\widetilde{\omega}}$. The left-hand side of the equation represents the wave equation for the electric field, while the first term on the right-hand side describes the contribution of the source charge density to the electric field; the second term on the right-hand side involves the divergence of the vorticity field, which represents the tendency of the vorticity to cause stretching and rotation of the electric field lines; the third term involves the gradient of the vorticity divergence, which represents the effect of the vorticity on the scalar potential of the electric field; the next two terms involve the dot product of the electric field with the gradient and the vorticity, respectively, and describe the interaction between the electric field and the vorticity; the sixth term involves the cross-product between the electric field and the curl of the vorticity and describes the tendency of the vorticity to induce circulation in the electric field. Finally, the last term involves the cross-product between the vorticity and the curl of the electric field and describes the tendency of the electric field to induce a rotation in the vorticity field. Overall, this equation describes the complex interplay between the electric field and the vorticity field and provides insight into the behavior of electromagnetic waves in various physical systems. The resulting Eq.~\ref{eq11a} is complex to analyse but we will consider the first and last term on the r.h.s., giving
\begin{equation}~\label{E13}
\nabla^2\vb{E} - \frac{1}{c^2} \frac{\partial^2 \vb{E}}{\partial t^2} = -\frac{1}{\epsilon_0}\nabla \rho - \left[\pmb{\widetilde{\omega}} \times \frac{\partial \pmb{B}}{\partial t} \right].
\end{equation}
Let's consider each term: i) $\frac{1}{\epsilon_0}\nabla \rho$, this term is classical, associated to Langmuir waves, or electron plasma waves, for example; ii) $\left[\pmb{\widetilde{\omega}} \times \frac{\partial \pmb{B}}{\partial t} \right]$ represents an interaction between a vector field (potentially a velocity or rotational field) and the time derivative of the magnetic field. It represents a new term force.

\begin{figure}
    \centering
    \includegraphics[width=0.48\textwidth]{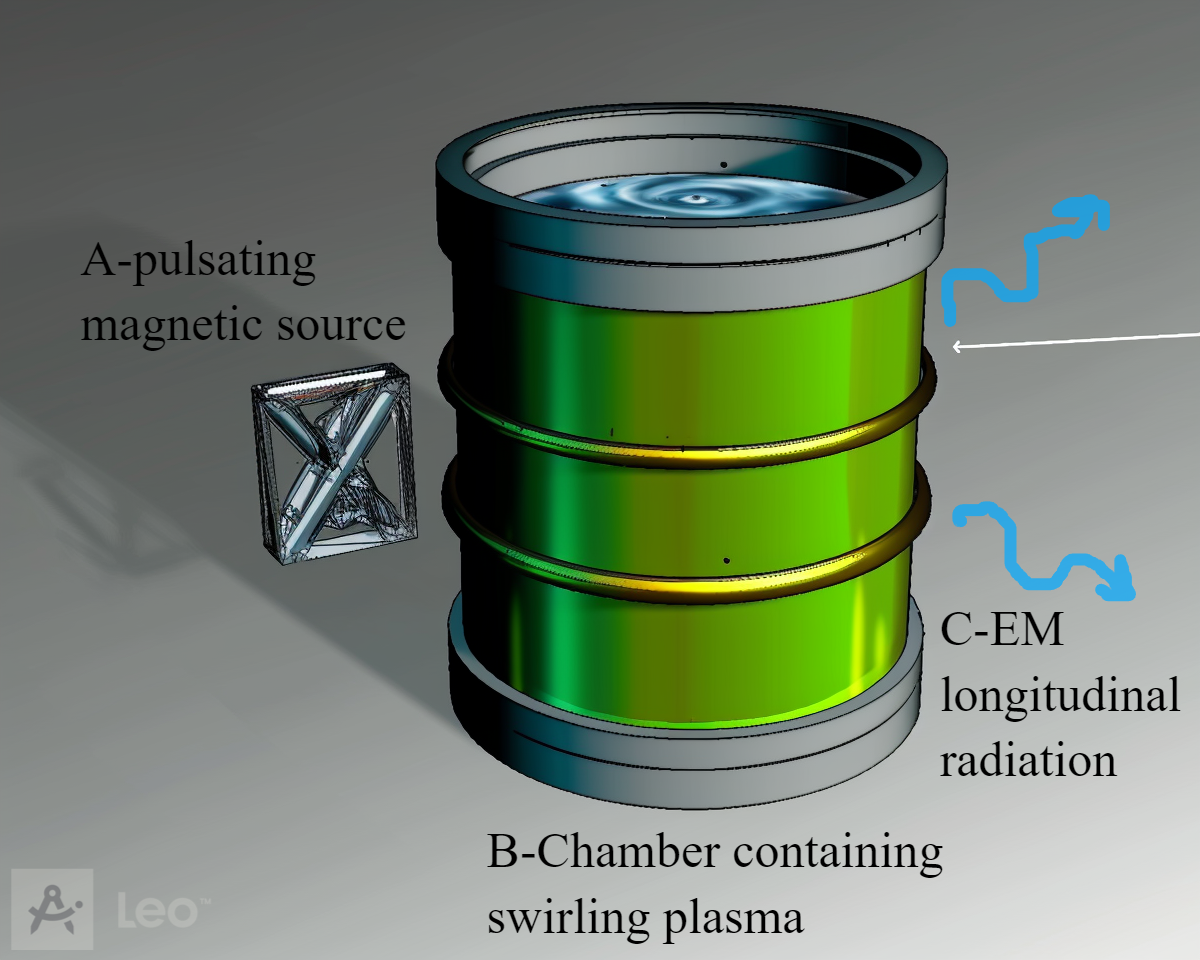}
    \caption{Mockup of an experimental apparatus for generating electromagnetic longitudinal waves. Component A represents the pulsating magnetic field generator. Component B is the cylindrical chamber containing swirling plasma, crucial for the wave generation process. Component C shows the resultant electromagnetic (EM) longitudinal waves (indicated by blue arrows), which are generated as a result of the cross product interaction between the magnetic field and the swirling plasma. Figure created using https://www.getleo.ai/.}
    \label{fig:swirl}
\end{figure}

Based on the interaction between rotating electromechanical systems, and temporally varying magnetic fields, we are able to suggest an experimental setup characterized by a dual-modality operational framework. At the core of this device is either a lattice designed with interlaced wire medium metamaterial (possible longitudinal EM wave generator)~\cite{Sakhno} or a rotational plasma arrangement (some methodologies are discussed in Refs.~\cite{Chang,Zhang}). These rotational systems are aligned such that their vectorial representations are orthogonal to the vector field of an externally imposed, pulsed magnetic field, thereby optimizing the conditions for maximal cross-product interaction. Specifically, the metamaterial lattice is designed to exhibit spatial anisotropy due to its cubic lattices and the way these lattices are interlaced manipulate the wave vector and dielectric permittivity characteristics to support longitudinal wave propagation, while the plasma setup facilitates angular momentum in ionized gases~\cite{Chang,Wouter,Durrani,Yu}. The pulsating magnetic field, applied in quadrature to these rotational vectors, serves as a dynamic perturbation, poised to elicit a rich spectrum of wave phenomena, both longitudinal and transverse. An illustrative example of possible experiment is proposed in Fig.~\ref{fig:swirl}.

A self-propelling device incorporating the previously mentioned principle and producing thrust with a cylindrical setup of radius 3.0 metres and height 1.0 metres (dimensions chosen for illustrative purpose) can be proposed based on the last term of force. It leverages the synergy of electric and magnetic fields, with a focus on a low-level charge density $\rho(z) =0.01$ C$/$m$^3$ (with Z the vertical axis), an angular frequency $\omega(z)$ of the order of 10 kHz (as, for example, observed in Hall Thrusters, see also Ref.~\cite{Oreshko}) and a dynamic magnetic field $dB/dt(z)$ (with applied frequency source of the order of MHz). Observe that the cross product $\left[\pmb{\widetilde{\omega}} \times \frac{\partial \pmb{B}}{\partial t} \right]$ determines the direction of motion. This approach, based on integrating the differential equation for the electric field term of force $\rho(z) E_z(z, t)$ across the cylinder's volume, suggests the possibility to generate thrust capability exceeding 2000 Newtons (the computational model and code detailing this concept are available in GitHub repository~\cite{Github2}.)

\subsection{Generation of electron beams carrying orbital angular momentum}

The generation of electron beams carrying orbital angular momentum (OAM)~\cite{Sun2023, Saitoh2022, Yan2022, Pastor2022, Tavabi2021, Liu2020, Xu2019, Nguyen2020, Eickhoff2020}
refers to the creation of electron beams that possess a well-defined helical phase front, similar to a spiral staircase. This helical structure of the electron beam arises due to the presence of OAM, which is a property of a wave that describes the rotation of the wave's phase around its axis of propagation.

One way to generate electron beams carrying OAM is by using electron holography, which is a technique that allows for the measurement and manipulation of electron wavefronts~\cite{Mafakheri2017, Grillo2016arXiv, Grillo2015PRL, Shi2023, Grillo2017}. In electron holography, a beam of electrons is split into two paths, with one path being used as a reference wave and the other passing through a sample. The two beams are then recombined, resulting in an interference pattern that contains information about the electron wavefront. By manipulating the electron wavefront using holography, it is possible to introduce a helical phase shift into the electron beam, thereby generating an electron beam carrying OAM. Such electron beams have been used in a variety of applications, such as electron microscopy and nanofabrication, and have the potential to revolutionize fields such as quantum computing and communication. We will illustrate how this formalism can be used to this issue.

The phase shift associated with an electron beam carrying OAM can be described mathematically using the following equation:
\begin{equation}~\label{eq13}
    \Psi(r,\theta, z) = A(r, \theta, z)\exp(il \theta),
\end{equation}
where $\Psi$ is the electron wavefunction, $A$ is the amplitude of the wavefunction, $r$, $\theta$, and $z$ are the cylindrical coordinates, and $l$ is the azimuthal quantum number that determines the amount of OAM carried by the electron beam. The total angular momentum $J$ of an electron beam carrying OAM can be expressed as 
$J = \hbar l$, where $\hbar$ is the reduced Planck constant. We may notice that the axial vector $\pmb{\widetilde{\omega}}$ has a dimension in $1/L$ and this fact suggests we postulate that the axial vector can be expressed as a spherical harmonic function $Y_{lm}(\theta,\phi)$ times a radial vector $\vb{r}/r$, where $\vb{r}$ is the radial unit vector~\cite{Allen1992, Lichte2008, McMorran2011, Bliokh2017}.

Based on this postulate, it was derived expressions for the divergence of the vector $\Phi(r) \;\pmb{\widetilde{\omega}}$ using the properties of spherical harmonics. Specifically, it is used the expressions for the divergence of $\Phi(r) \vb{Y}_{l,l+1,m}(\theta,\phi)$, $\Phi(r) \vb{Y}_{l,l,m}(\theta,\phi)$, and $\Phi(r) \vb{Y}_{l,l-1,m}(\theta,\phi)$, which are derived using the differential operators for spherical coordinates. These expressions involve derivatives of the radial function $\Phi(r)$ and the spherical harmonic functions, and they provide a way to relate the divergence of $\Phi(r) \;\pmb{\widetilde{\omega}}$ to the function $\Phi(r)$ and its derivatives (see also Ref.~\cite{Edmonds} with further information on the properties and use of spherical harmonics in physics and see also, Refs.~\cite{Uchida,Verbeeck,Bliokh} for applications). Hence, the Axial vector $\pmb{\widetilde{\omega}}$ is defined by
\begin{equation}~\label{eq14}
\pmb{\widetilde{\omega}}=\frac{\vb{r}}{r} Y_{lm}(\theta,\phi).
\end{equation}

A more extended expression for $\pmb{\widetilde{\omega}}$ in terms of vector spherical harmonics is
\begin{equation}~\label{eq15}
\pmb{\widetilde{\omega}}=-\left[\frac{l+1}{2l+1}\right]^{1/2} \vb{Y}{l,l+1,m} +\left[\frac{l}{2l+1}\right]^{1/2} \vb{Y}{l,l-1,m}
\end{equation}
and for the divergence of $\Phi(r) \, \pmb{\widetilde{\omega}}$we may write:
\begin{equation}~\label{eq16}
\begin{aligned}
\pmb{\nabla} \cdot [\Phi(r) \, \pmb{\widetilde{\omega}}] &= -\left[ \frac{l+1}{2l+1} \right]^{1/2} \pmb{\nabla} \cdot [\Phi(r) Y_{l,l-1,m}(\theta,\phi)] \\
&\qquad + \left[ \frac{l}{2l+1} \right]^{1/2} \pmb{\nabla} \cdot [\Phi(r) \vb{Y}_{l,l-1,m}(\theta,\phi)].
\end{aligned}
\end{equation}
Since the equations discussed thus far are based on the mathematics of vector spherical harmonics and the characteristics of angular momentum, they offer an effective foundation for explaining helical waves and OAM in electron beams.
However, it is important to note that these equations are not necessarily an improvement over existing methods for describing OAM in electromagnetic waves. Instead, they offer an alternative framework for understanding and analyzing electron beams carrying OAM, especially in the context of electron holography and other experimental techniques, but they can be useful for certain applications and may help researchers gain new insights into the behavior of electron beams with OAM. For instance, in electron microscopy~\cite{McMorran2011}, these equations can potentially lead to an improvement in the analysis of the electron beam's interaction with samples: when an electron beam with OAM interacts with a sample, the helical phase front can be influenced by the sample's structure and composition, leading to a change in the OAM distribution~\cite{Bliokh2017}.

By using the derived equations to describe the helical phase front and the divergence of the axial vector, we can gain a deeper understanding of the electron beam's interaction with the sample, helping to decipher the effects of the sample's structure on the electron beam's OAM distribution and ultimately lead to a more accurate interpretation of the electron microscopy images. In addition, these equations can potentially be utilized to design new electron holography experiments or techniques that take advantage of the OAM properties of electron beams, for example, by manipulating the OAM of an electron beam, new methods to probe specific structural features of a sample or improve the resolution of electron microscopy can be devised~\cite{Shangguan2023, Paroli2023, Suciu2023}.

For instance, consider an application in which researchers are working with twisted light beams (optical vortices) carrying OAM in optical communication systems~\cite{Shangguan2023, Paroli2023, Suciu2023, Luo2022, Xie2023, Kovalev2021, Pryamikov2021, Innes2019, Zhao2020}. They could use these equations to analyze the effects of different media on the propagation of the twisted light beams and their OAM properties. By understanding the changes in the axial vector and its divergence, researchers could optimize the design of optical communication systems to minimize the loss of OAM information during transmission, thus improving the efficiency and reliability of these systems. Similarly, these equations can be applied in the field of plasmonics, where the interaction of light with metallic nanostructures to confine and manipulate EM waves on the nanoscale is of importance~\cite{Chau2023, Jia2023, KumarSingh2021}. By using these equations, we can analyze the impact of different nanostructures on the OAM properties of the incident EM waves and design nanostructures that can efficiently manipulate the OAM properties for various applications, such as optical tweezers, super-resolution imaging, or quantum information processing~\cite{Fisicaro_2016,Chicone_2012,Anderson 2008,Tsoi,Trier}.

\subsection{Application of the axial vector and its divergence to plasmonics}

To illustrate the potential use of your equations for the axial vector and its divergence, let's consider an example in the field of plasmonics. We'll analyze the interaction of a twisted light beam with a metallic nanostructure, such as a metallic nanorod, and see how the nanostructure affects the OAM properties of the incident EM wave.

First, we will use the axial vector $\pmb{\widetilde{\omega}}$, Eq.~\ref{eq14} defined before. Next, we'll use the expression for the divergence of $\Phi(r) \, \pmb{\widetilde{\omega}}$, Eq.~\ref{eq16}.

\begin{figure}
    \centering
    \includegraphics[width=0.52\textwidth]{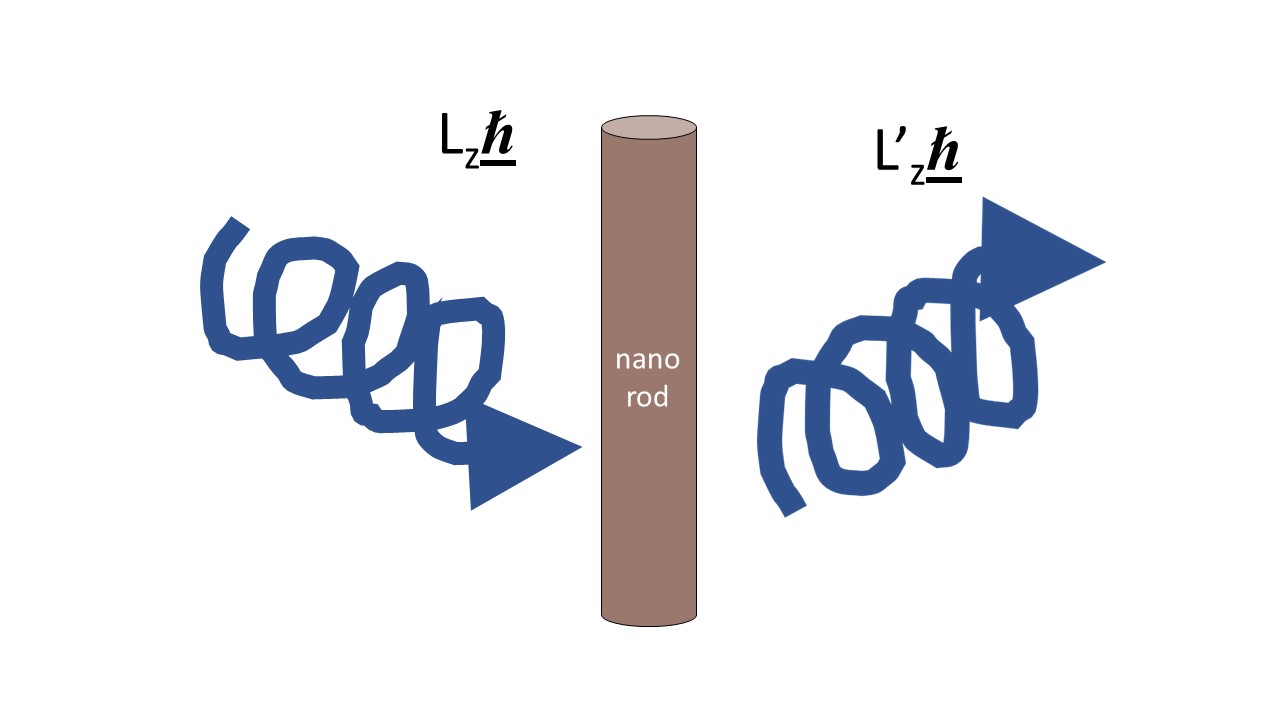}
    \caption{Illustration of the interaction between a twisted light beam and a metallic nanorod, demonstrating the principles of electron holography in generating electron beams with orbital angular momentum (OAM). The figure depicts the helical phase front of the electron beam, induced by the OAM, as it interacts with the nanorod, highlighting the potential for manipulation and analysis of the OAM properties in various applications, including electron microscopy and optical communications.}
    \label{fig:twisted}
\end{figure}

We'll consider the interaction between the twisted light beam and the metallic nanorod, the situation illustrated in Fig.~\ref{fig:twisted}.

This interaction modifies the amplitude $\Phi(r)$ of the EM wave. In this case, we can calculate the new amplitude $\Phi'(r)$ after the interaction with the nanorod by solving Maxwell's equations or using numerical methods like finite-difference time-domain (FDTD) simulations. With the new amplitude $\Phi'(r)$, the new axial vector $\pmb{\widetilde{\omega}'}$ can be computed and as well as its divergence. By comparing the divergence of $\pmb{\widetilde{\omega}}$ and $\pmb{\widetilde{\omega}'}$, we can analyze the impact of the metallic nanorod on the OAM properties of the incident EM wave.

\begin{figure}
    \centering
    \includegraphics[width=0.45\textwidth]{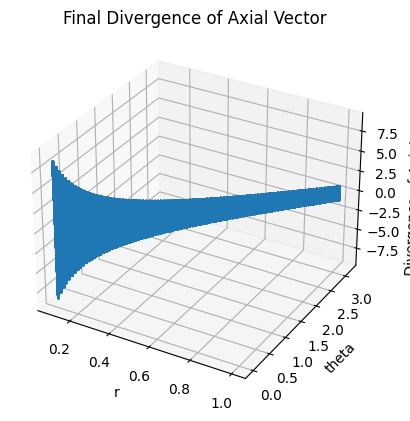}
    \caption{Graphical representation of the calculated differences in the axial vector \(\pmb{\widetilde{\omega}}\), as derived from the proposed method. This figure quantitatively illustrates the variations in the axial vector resulting from specific interactions or conditions in the study, emphasizing the method's capability to detect and analyze subtle changes in electron beam properties in plasmonic and other advanced applications.}
    \label{fig:met1}
\end{figure}

We will develop further now a simple model using the axial vector and its divergence to understand the impact of a metallic nanorod on the OAM properties of an incident EM wave. We'll analyze the changes in the axial vector and its divergence before and after interaction with the nanorod. Let's assume the following: The incident twisted light beam has a helical wavefront described by the wave function $\Psi(r,\theta,z) = A(r,\theta,z)\exp(il\theta)$, where $l$ is the azimuthal quantum number determining the OAM. The metallic nanorod has a length $L$ and a radius $a$. It is positioned along the z-axis and interacts with the incident EM wave (see Fig.~\ref{fig:twisted}).

\begin{itemize}

\item Step 1: Calculating the initial axial vector and its divergence for the incident EM wave. Using the equations, we can compute the initial axial vector $\pmb{\widetilde{\omega}}$ and its divergence $\pmb{\nabla} \cdot (\Phi(r) \; \pmb{\widetilde{\omega}})$ before the interaction with the nanorod.

\item Step 2: Defining the interaction between the twisted light beam and the metallic nanorod. For simplicity, we'll assume that the interaction between the twisted light beam and the nanorod can be represented by an interaction factor $\alpha$. This factor depends on the size, shape, and material properties of the nanorod, as well as the wavelength of the incident EM wave. The new amplitude after the interaction can be represented as $\Phi'(r) = \alpha \Phi(r)$.

\item Step 3: Calculating the new axial vector and its divergence after the interaction. Using the new amplitude $\Phi'(r)$, we can compute the new axial vector $\pmb{\widetilde{\omega}'}$ and its divergence $\pmb{\nabla} \cdot (\Phi'(r) \; \pmb{\widetilde{\omega}'})$ after the interaction with the nanorod.

\item Step 4: Comparing the initial and final axial vectors and their divergences. By comparing the axial vectors $\pmb{\widetilde{\omega}}$ and $\pmb{\widetilde{\omega}'}$, as well as their divergences, we can analyze the changes in the OAM properties of the incident EM wave due to the interaction with the metallic nanorod.

\item Step 5: Optimizing the design of the metallic nanorod.
\end{itemize}
Based on the analysis of the changes in the axial vector and its divergence, researchers can optimize the design parameters of the metallic nanorod (such as length, radius, and material) to efficiently manipulate the OAM properties of the incident EM waves for specific applications. This procedure is illustrated in Fig.~\ref{fig:met1}, and the code used for the implementation can be found on Github at~\cite{github}. 

This simple model helps illustrate the use of the axial vector and its divergence in analyzing the impact of a metallic nanorod on the OAM properties of an incident EM wave. It also provides a starting point for designing nanostructures that can efficiently manipulate the OAM properties for various applications. However, a more realistic model would require solving Maxwell's equations or using numerical methods like FDTD simulations to account for the complex interaction between the twisted light beam and the nanostructure.


\section{Conclusion}

The modified Poisson equation provides insights into several physical systems. It is developed from variational concepts and its potential in spintronics and its application in modelling resonant structures and nonlocal gravity theories enable new research directions, including new sources of propulsion, energy conversion and communication, although without any doubt future research is still needed to fully understand the potential of these applications. Through the use of the axial vector and its divergence when conducting plasmonics research, a novel approach was proposed to control the orbital angular momentum of electromagnetic waves as they interact with nanostructured surfaces. It is expected interesting ramifications for metamaterials and nonlinear optics, particularly with regard to producing and managing longitudinal electromagnetic waves. The potential for new wave induction techniques is highlighted by the new components of the Poisson's equation, in particular the interaction between rotational vector fields and time-varying magnetic fields. Our suggested experimental setup seeks to investigate these interactions further, with the potential to transform data storage and communication technologies by merging spinning plasma settings or metamaterial lattices.

\begin{acknowledgments}
This research was conducted autonomously, without reliance on external financial aid or assistance.
\end{acknowledgments}

\bibliographystyle{unsrt}

\begin{thebibliography}{199}

\bibitem{Poisson} Poisson, C. (1811). Mécanique analytique. Paris.

\bibitem{Maxwell} Maxwell, J. C. (1873). A Treatise on Electricity and Magnetism. Oxford.

\bibitem{Pinheiro2013} Pinheiro, M. J. (2013). A Variational Method in Out-of-Equilibrium Physical Systems. Sci Rep, 3, 3454. \url{https://doi.org/10.1038/srep03454}

\bibitem{Myers1998} Myers, W. D., $\&$ Swiatecki, W. J. (1998). The nuclear shape transition and its relation to a new class of mesoscopic systems. Physics Letters B, 441(1-4).

\bibitem{Wright2001} Wright, E. L. (2001). The intergalactic medium. In M. S. Dresselhaus $\&$ G. Dresselhaus (Eds.), Universe in a Nutshell: The Physics of Everything (pp. 49-57). New York, NY: Springer.

\bibitem{Zutic2004} Zutic, I., Fabian, J., $\&$ Das Sarma, S. (2004). Spintronics: Fundamentals and applications. Reviews of Modern Physics, 76(2), 323.

\bibitem{Serdyukov2001} Serdyukov, V., Semchenko, I., Tretyakov, S., $\&$ Sihvola, A. (2001). Electromagnetics of Bi-anisotropic Materials: Theory and Applications. Amsterdam, Netherlands: Gordon and Breach.

\bibitem{Gilbarg1983} Gilbarg, D., $\&$ Trudinger, N. S. (1983). Elliptic Partial Differential Equations of Second Order (2nd ed.). Berlin, Germany: Springer.

\bibitem{Reddy2006} Reddy, J. N. (2006). An Introduction to the Finite Element Method (3rd ed.). New York, NY: McGraw-Hill.

\bibitem{Mammoli2018} Mammoli, A. A., $\&$ Kassab, Y. M. (2018). Boundary Element Analysis: Mathematical Aspects and Applications. Boca Raton, FL: CRC Press.

\bibitem{Trefethen1997} Trefethen, L. N. (1997). Spectral methods. In P. G. Ciarlet $\&$ J.-L. Lions (Eds.), Handbook of Numerical Analysis (pp. 209-305). Amsterdam, Netherlands: Elsevier.

\bibitem{Tarassov2022} Tarasov, V.E. (2022). Nonlocal classical theory of gravity: massiveness of nonlocality and mass shielding by nonlocality. Eur. Phys. J. Plus, 137, 1336. \url{https://doi.org/10.1140/epjp/s13360-022-03512-x}

\bibitem{Mashhoon2022} Mashhoon, B. (2022). Nonlocal Gravity. Oxford Academic. \url{https://doi.org/10.1093/oso/9780198803805.001.0001}, accessed 18 Jan. 2024

\bibitem{Duadi} Pawar, S., Duadi, H., $\&$ Fixler, D. (2023). Recent Advances in the Spintronic Application of Carbon-Based Nanomaterials. Nanomaterials, 13(3), 598-598. \url{https://doi.org/10.3390/nano13030598}

\bibitem{Lu} Lu, W.-t., $\&$ Yuan, Z. (2022). Progress in ultrafast spintronics research. Sci. Sin. Phys. Mech. Astron. 52(7), 270007-270007. \url{https://doi.org/10.1360/sspma-2021-0350}

\bibitem{Dei} Dey, C., Yari, P., $\&$ Wu, K. (2023). Recent advances in magnetoresistance biosensors: a short review. Nano Futures, 7(1), 012002-012002. \url{https://doi.org/10.1088/2399-1984/acbcb5}

\bibitem{Jonatan} Lenells, J. (2008). Poisson structure of a modified Hunter-Saxton equation. Journal of Physics A, 41(28), 285207-. \url{https://doi.org/10.1088/1751-8113/41/28/285207}

\bibitem{Ludovic} Burgnies, L., Vanbésien, O., Sadaune, V., Lippens, D., Nagle, J., $\&$ Vinter, B. (1994). Resonant tunneling structures with local potential perturbations. Journal of Applied Physics, 75(9), 4527-4532. \url{https://doi.org/10.1063/1.355945}

\bibitem{Joseph} Krasil’shchik, J., Verbovetsky, A., $\&$ Vitolo, R. (2017). Variational Poisson Structures. pp. 193-214. \url{https://doi.org/10.1007/978-3-319-71655-8_10}

\bibitem{Pinheiro2016} Pinheiro, M. J. (2016). Some effects of topological torsion currents on spacecraft dynamics and the flyby anomaly. Mon. Not. R. Astron. Soc. 461(4), 3948–3953. \url{https://doi.org/10.1093/mnras/stw1581}

\bibitem{Pinheiro2022} Pinheiro, M. J. (2022). Ergontropic Dynamics: Contribution for an Extended Particle Dynamics. In Bandyopadhyay, A., $\&$ Ray, K. (Eds.), Rhythmic Advantages in Big Data and Machine Learning. Springer. Singapore. \url{https://doi.org/10.1007/978-981-16-5723-8_3}

\bibitem{dynamics_and_transport}
Y. Feng, S. Lu, K. Wang, W. Lin, and D. Huang, ``Dynamics and transport of magnetized two-dimensional Yukawa liquids,'' 2019.

\bibitem{isomorph_invariance}
F. Lucco Castello, P. Tolias, J. Schmidt Hansen, and J. C. Dyre, ``Isomorph invariance and thermodynamics of repulsive dense bi-Yukawa one-component plasmas,'' Physics of Plasmas, 2019.

\bibitem{molecular_dynamics_study}
S. Maity and A. Das, ``Molecular dynamics study of crystal formation and structural phase transition in Yukawa system for dusty plasma medium,'' Physics of Plasmas, 2019.

\bibitem{Zaripov} Zaripov, F. (2020). Dark Matter as a Result of Field Oscillations in the Modified Theory of Induced Gravity. Symmetry, 12(1), 41. \url{https://doi.org/10.3390/sym12010041}

\bibitem{Hayashi} Eric Hayashi, Julio F. Navarro, and Volker Springel, MNRAS (2007) 377 (1) 50-62

\bibitem{intro_spintronics} Atsufumi Hirohata, Keisuke Yamada, Yoshinobu Nakatani, Ioan-Lucian Prejbeanu, Bernard Diény, Philipp Pirro, Burkard Hillebrands, ``Review on spintronics: Principles and device applications,''
J. Magn. Magn. Mater {\bf 509}, pp. 166711 (2020)
\url{https://doi.org/10.1016/j.jmmm.2020.166711}

\bibitem{Cheng} Shi, Cheng., Wen-Jeng, Hsueh. (2023). High spin current density in gate-tunable spin-valves based on graphene nanoribbons. Dental science reports, 13(1) \url{doi: 10.1038/s41598-023-36478-6}

\bibitem{Qi} Qi, An., Sébastien, Le, Beux., Ian, O'Connor., Jacques-Olivier, Klein. (2018). Large scale, high density integration of all spin logic.  131-136. \url{doi: 10.23919/DATE.2018.8341992}

\bibitem{Angizi} Shaahin, Angizi., Zhezhi, He., Yu, Bai., Jie, Han., Mingjie, Lin., Ronald, F., DeMara., Deliang, Fan. (2018). Leveraging Spintronic Devices for Efficient Approximate Logic and Stochastic Neural Networks.  397-402. \url{doi: 10.1145/3194554.3194618}

\bibitem{Jonietz} F., Jonietz., Sebastian, Mühlbauer., Christian, Pfleiderer., A., Neubauer., W., Münzer., Andreas, Bauer., T., Adams., Robert, Georgii., Peter, Böni., Rembert, A., Duine., K., Everschor., Markus, Garst., Achim, Rosch. (2010). Spin Transfer Torques in MnSi at Ultralow Current Densities. Science, 330(6011):1648-1651. \url{doi: 10.1126/SCIENCE.1195709}

\bibitem{Cockburn} Bruce, F., Cockburn. (2003). The emergence of high-density semiconductor-compatible spintronic memory.  321-326. \url{doi: 10.1109/ICMENS.2003.1222018}

\bibitem{Waser_2001} K. J. van Vlaenderen and A. Waser, {\it Hadronic Journal} {\bf 24}, 609 (2001)

\bibitem{Vlaenderen} K. J. van Vlaenderen, {\it A generalization of classical electrodynamics for the prediction of scalar field effects}, arXiv:physics/0305098v1 [physics.class-ph]

\bibitem{Khvorostenko_2002} N. P. Khvorostenko, {\it Izvestya vuzov, Fizika} {\bf 3}, 24 (2002)

\bibitem{Wesley_2002} C. Mostein and J. P. Wesley, {\it Europhysics Lett.} {\bf 59}, 514 (2002)

\bibitem{Podgany_2011} O. A. Zaymidoroga and D. V. Podgainy, {\it Observation of electroscalar Radiation during a Solar Eclipse}, p.84 in Cosmic Rays for Particle and Astroparticle Physics, Proc. of the 12th ICATPP Conference, (World Scientific, Singapore, 2011)

\bibitem{Hively} Hively, L. (2016). US 9,306,527 B1.

\bibitem{Jefimenko} Jefimenko, O. D. (1992). Causality, electromagnetic potentials, and longitudinal waves. American Journal of Physics, 60(9), 839-844.

\bibitem{Kong} Kong, J. A. (1998). Electromagnetic wave theory of closed structures. Progress in Electromagnetic Research, 18, 269-346.

\bibitem{Lindell} Lindell, I. V., $\&$ Sihvola, A. (2005). Perfect electromagnetic conductor. Journal of Electromagnetic Waves and Applications, 19(7), 861-869.

\bibitem{He} He, Q., Zhao, X., $\&$ Mu, Q. (2006). Electromagnetic wave in a plasma-filled coaxial cable. IEEE Transactions on Plasma Science, 34(5), 1778-1781.

\bibitem{Shukla} Shukla, P. K., $\&$ Eliasson, B. (2007). Electromagnetic waves in dusty plasmas. Physics of Plasmas, 14(5), 054502.

\bibitem{Wu} Wu, H., Li, Q., $\&$ Wu, J. (2012). Electromagnetic wave propagation in graphene. Journal of Physics: Condensed Matter, 24(16), 164206.

\bibitem{Sakhno} Sakhno, D., Koreshin, E., $\&$ Belov, P. A. (2021). Longitudinal electromagnetic waves with extremely short wavelength. Phys. Rev. B, 104(10), L100304. \url{https://doi.org/10.1103/PhysRevB.104.L100304}


\bibitem{Chang} Chang, C. C., $\&$ Lundgren, T. S. (1959). Flow of an Incompressible Fluid in a Hydromagnetic Capacitor. Phys. Fluids, 2(6), 627–632. \url{https://doi.org/10.1063/1.1705964}

\bibitem{Zhang} Zhang, K., Wang, Y., Tang, H., Li, Y., Wang, B., York, T. M., Yang, L. (2020). Two-dimensional analytical investigation into energy conversion and efficiency maximization of magnetohydrodynamic swirling flow actuators. Energy, 209, 118479. \url{https://doi.org/10.1016/j.energy.2020.118479}

\bibitem{Wouter} Bos, W. J. T., Neffaa, S., $\&$ Schneider, K. (2008). Rapid generation of angular momentum in bounded magnetized plasma. Physical Review Letters, 101(23), 235003. \url{https://doi.org/10.1103/PHYSREVLETT.101.235003}

\bibitem{Durrani} Durrani, I. R. (2012). Photon orbital angular momentum in a plasma vortex. Bulletin of Pure $\&$ Applied Sciences-Physics.

\bibitem{Yu} Fitzpatrick, R., $\&$ Yu, E. (1997). Angular momentum injection into a Penning–Malmberg trap. Phys. Plasmas, 4(4), 917-930. \url{https://doi.org/10.1063/1.872208}

\bibitem{Oreshko} Oreshko. (2001). Generation of Strong Fields in Plasma. Doklady Physics, 46(1), 9–11. (Original work published in Doklady Akademii Nauk, Vol. 376, No. 2, 2001, pp. 183–185)

\bibitem{Github2} Pinheiro, M. J. (2024). CylindricalPlasmaThruster.ipynb. GitHub repository. \url{https://github.com/mjgpinheiro/Physics_models/blob/main/CylindricalPlasmaThruster.ipynb}

\bibitem{Sun2023} Sun, H., Liu, B., $\&$ Feng, C. (2023). Short-wavelength radiation pulses with time-varying orbital angular momentum from tailored relativistic electron beams. Opt. Lett. \url{https://doi.org/10.1364/ol.496317}

\bibitem{Saitoh2022} Saitoh, K., Yonezawa, T., Nambu, H., Tanimura, S., $\&$ Uchida, M. (2022). Orbital Angular Momentum Resolved Convergent-Beam Electron Diffraction by the Post-Selected Injection of Electron Beam. Microscopy. \url{https://doi.org/10.1093/jmicro/dfac046}

\bibitem{Yan2022} Yan, J.-A., $\&$ Geloni, G. (2022). Self-seeded free-electron lasers with orbital angular momentum. Adv. Photonics Nexus. \url{https://doi.org/10.1117/1.APN.2.3.036001}

\bibitem{Pastor2022} Pastor, I., Alvarez-Estrada, R. F., Roso, L., $\&$ Castejón, F. (2022). Fundamental Studies on Electron Dynamics in Exact Paraxial Beams with Angular Momentum. Photonics. \url{https://doi.org/10.3390/photonics9100693}

\bibitem{Tavabi2021} Tavabi, A. H., Rosi, P., Rotunno, E., Roncaglia, A., Belsito, L., Frabboni, S., Pozzi, G., Gazzadi, G. C., Lu, P.-H., Nijland, R., Ghosh, M., Tiemeijer, P., Karimi, E., Dunin-Borkowski, R. E., $\&$ Grillo, V. (2021). Experimental Demonstration of an Electrostatic Orbital Angular Momentum Sorter for Electron Beams. Phys. Rev. Lett., 126(9), 094802. \url{https://doi.org/10.1103/PHYSREVLETT.126.094802}


\bibitem{Liu2020} Liu, P., Yan, J., Afanasev, A., Benson, S. V., Hao, H., Mikhailov, S., Popov, V., $\&$ Wu, Y. (2020). Orbital angular momentum beam generation using a free-electron laser oscillator. arXiv: Accelerator Physics. \url{https://arxiv.org/pdf/2007.15723.pdf}

\bibitem{Xu2019} Xu, P., $\&$ Zhang, C. (2019). Orbital angular momentum microwave generated by free electron beam. In International Conference on Communications. \url{https://doi.org/10.1007/978-3-030-41114-5_14}

\bibitem{Nguyen2020} Nguyen, K. X., Jiang, Y., Cao, M. C., Purohit, P., Yadav, A. K., García-Fernández, P., Tate, M. W., Chang, C. S., Aguado-Puente, P., Íñiguez, J., Gómez-Ortiz, F., Gruner, S. M., Junquera, J., Martin, L. W., Ramesh, R., $\&$ Muller, D. A. (2020). Transferring Orbital Angular Momentum to an Electron Beam Reveals Toroidal and Chiral Order. arXiv: Materials Science. \url{https://arxiv.org/pdf/2012.04134}

\bibitem{Eickhoff2020} Eickhoff, K., Rathje, C., Köhnke, D., Kerbstadt, S., Englert, L., Bayer, T., Schäfer, S., Wollenhaupt, M. (2020). Orbital angular momentum superposition states in transmission electron microscopy and bichromatic multiphoton ionization. New J. Phys. \url{https://doi.org/10.1088/1367-2630/ABBE54}

\bibitem{Mafakheri2017} Mafakheri, E., Tavabi, A. H., Lu, P.-H., et al. (2017). Realization of electron vortices with large orbital angular momentum using miniature holograms fabricated by electron beam lithography. Appl. Phys. Lett., 110. \url{https://doi.org/10.1063/1.4977879}

\bibitem{Grillo2016arXiv} Grillo, V., Gazzadi, G. C., Mafakheri, E., et al. (2016). Realization of electron vortices with large orbital angular momentum using miniature holograms fabricated by electron beam lithography. arXiv: Quantum Physics. \url{https://arxiv.org/abs/1612.00654}

\bibitem{Grillo2015PRL} Grillo, V., Gazzadi, G. C., Mafakheri, E., et al. (2015). Holographic Generation of Highly Twisted Electron Beams. Phys. Rev. Lett., 114. \url{https://doi.org/10.1103/PhysRevLett.114.034801}


\bibitem{Shi2023} Shi, Z., Wan, Z., Zhan, Z., et al. (2023). Super-resolution orbital angular momentum holography. Nature Communications, 14. \url{https://doi.org/10.1038/s41467-023-37594-7}

\bibitem{Grillo2017} Grillo, V., Tavabi, A. H., Venturi, F., et al. (2017). Measuring the orbital angular momentum spectrum of an electron beam. Nature Communications, 8. \url{https://doi.org/10.1038/NCOMMS15536}

\bibitem{Allen1992} Allen, L., Beijersbergen, M. W., Spreeuw, R. J. C., $\&$ Woerdman, J. P. (1992). Orbital angular momentum of light and the transformation of Laguerre-Gaussian laser modes. Phys. Rev. A, 45, 8185. \url{https://doi.org/10.1103/PhysRevA.45.8185}

\bibitem{Lichte2008} Lichte, H., $\&$ Lehmann, M. (2008). Electron holography – basics and applications. Reports on Progress in Physics, 71, 016102. \url{https://doi.org/10.1088/0034-4885/71/1/016102}


\bibitem{McMorran2011} McMorran, B. J., et al. (2011). Electron Vortex Beams with High Quanta of Orbital Angular Momentum. Science, 331, 192. \url{https://doi.org/10.1126/science.1198804}

\bibitem{Bliokh2017} Bliokh, K. Y., et al. (2017). Theory and applications of free-electron vortex states. Physics Reports, 690, 1-70. \url{https://doi.org/10.1016/j.physrep.2017.05.006}

\bibitem{Shangguan2023} Shangguan, J.-t., Sun, Q., Jin, L., et al. (2023). The coupling of multi-channel optical vortices based on angular momentum conservation using a single-layer metal metasurface. EPL (Europhysics Letters). \url{https://doi.org/10.1209/0295-5075/acb2f4}

\bibitem{Paroli2023} Paroli, B., Siano, M., Cremonesi, L., $\&$ Potenza, M. N. (2023). High data-transfer density using 4-states optical vortices for deep space optical communication links. International Conference on Optical Network Design and Modelling.

\bibitem{Suciu2023} Suciu, S., Bulzan, G. A., Isdraila, T.-A., et al. (2023). Quantum communication networks with optical vortices.

\bibitem{Edmonds} A. R. Edmonds, {\it Angular Momentum in Quantum Mechanics} (Princeton, NJ, 1974)

\bibitem{Uchida} Uchida, M., $\&$ Tonomura, A. (2010). Generation of electron beams carrying orbital angular momentum. Nature, 464(7289), 737-739.

\bibitem{Verbeeck} Verbeeck, J., Tian, H., $\&$ Schattschneider, P. (2010). Production and application of electron vortex beams. Nature, 467(7313), 301-304.

\bibitem{Bliokh} Bliokh, K. Y., Rodríguez-Fortuño, F. J., Nori, F., $\&$ Zayats, A. V. (2015). Spin–orbit interactions of light. Nature Photonics, 9(12), 796-808.

\bibitem{Shangguan2023} Shangguan, J.-t., Sun, Q., Jin, L., et al. (2023). The coupling of multi-channel optical vortices based on angular momentum conservation using a single-layer metal metasurface. EPL (Europhysics Letters). \url{https://doi.org/10.1209/0295-5075/acb2f4}

\bibitem{Paroli2023} Paroli, B., Siano, M., Cremonesi, L., $\&$ Potenza, M. N. (2023). High data-transfer density using 4-states optical vortices for deep space optical communication links. International Conference on Optical Network Design and Modelling.

\bibitem{Suciu2023} Suciu, S., Bulzan, G. A., Isdraila, T.-A., et al. (2023). Quantum communication networks with optical vortices.

\bibitem{Luo2022} Luo, H., Yang, K., Li, P.-W., et al. (2022). Generation and verification of optical vortices with controlled phase based on coherent beam combining. Physica Scripta. \url{https://doi.org/10.1088/1402-4896/ac91ff}

\bibitem{Xie2023} Xie, Z. (2023). Nondiffractive polarization feature of optical vortices. Advanced Photonics. \url{https://doi.org/10.1117/1.ap.5.3.030503}

\bibitem{Kovalev2021} Kovalev, A.A. (2021). Optical vortices with an infinite number of screw dislocations. Computer Optics. \url{https://doi.org/10.18287/2412-6179-CO-866}

\bibitem{Pryamikov2021} Pryamikov, A. D., Hadzievski, L., Fedoruk, M. P., et al. (2021). Optical vortices in waveguides with discrete and continuous rotational symmetry. Journal of the European Optical Society: Rapid Publications. \url{https://doi.org/10.1186/S41476-021-00168-5}

\bibitem{Innes2019} Innes, T., Elliott, O., $\&$ Scruggs, S. (2019). Optical networking with hybrid optical vortices. US Patent 10506312. \url{https://www.freepatentsonline.com/10506312.html}

\bibitem{Zhao2020} Zhao, L., Jiang, T., Mao, M., et al. (2020). Improve The Capacity Of Data Transmission In Orbital Angular Momentum Multiplexing By Adjusting Link Structure. IEEE Photonics Journal. \url{https://doi.org/10.1109/JPHOT.2020.2985728}

\bibitem{Chau2023} Chau, K. K. W. (2023). Multiscale models of plasmonic structural colors with nanoscale surface roughness. Optics Letters. \url{https://doi.org/10.1364/ol.474703}

\bibitem{Jia2023} Jia, H., Tsoi, C. C., El Abed, A. I., et al. (2023). Metallic Plasmonic Nanostructure Arrays for Enhanced Solar Photocatalysis. Laser $\&$ Photonics Reviews. \url{https://doi.org/10.1002/lpor.202200700}

\bibitem{KumarSingh2021} Singh, A. K., Kumar, A., Dixit, S., $\&$ Kumar, A. (2021). Interaction of Light with Plasmonic Nanostructures Fabricated by Nanosphere Lithography. \url{https://doi.org/10.1007/978-981-15-9259-1_189}

\bibitem{Fisicaro_2016} Fisicaro, G., Genovese, L., Andreussi, O., Marzari, N., $\&$ Goedecker, S. (2016). A generalized Poisson and Poisson-Boltzmann solver for electrostatic environments. Journal of Chemical Physics, 144, 014103.

\bibitem{Chicone_2012} Chicone, C., $\&$ Mashhoon, B. (2012). Nonlocal gravity: Modified Poisson's equation. Journal of Mathematical Physics, 53, 042501.

\bibitem{Anderson 2008} John D. Anderson, James K. Campbell, John E. Ekelund, Jordan Ellis, and James F. Jordan, Phys. Rev. Lett. {\bf 100}, 091102 (2008)

\bibitem{Tsoi} Tsoi, M., Jansen, A. G. M., Bass, J., and Chiang, W. H. (1998). Spin current in magnetic nanostructures. Journal of Magnetism and Magnetic Materials, 200(1-3), 479-490

\bibitem{Trier} Felix Trier, Diogo Vaz, Pierre Bruneel, Paul Noël, Albert Fert, et al.. Electric-Field Control of Spin Current Generation and Detection in Ferromagnet-Free SrTiO 3 -Based Nanodevices. Nano Letters, 2019, 20 (1), pp.395-401. ⟨10.1021/acs.nanolett.9b04079⟩. ⟨hal-02517872⟩

\bibitem{github} Pinheiro, M. J. (2021). Physics\_models: OAM\_EM.ipynb. GitHub repository. \url{https://github.com/mjgpinheiro/Physics_models/blob/main/OAM_EM.ipynb}

\end{thebibliography}

\end{document}